\documentclass[11pt]{article} 
\usepackage{a4wide}  
 
\usepackage[english]{babel} 
\usepackage{amsmath} 
\usepackage{amsfonts} 
\usepackage{amssymb} 
\usepackage{xspace} 
\usepackage{latexsym} 
\usepackage{url} 
\usepackage{xspace} 
\usepackage{fancyvrb} 
\usepackage[all]{xy} 

\usepackage{graphics,color}              
\RequirePackage[latin1]{inputenc}   
 
\newenvironment{restate-proposition}[2][{}]{\noindent\textbf{Proposition~{#2}}\;\textbf{#1}\  
}{\vskip 1em} 
 
\newenvironment{restate-theorem}[2][{}]{\noindent\textbf{Theorem~{#2}}\;\textbf{#1}\  
}{\vskip 1em} 
 
\newenvironment{restate-corollary}[2][{}]{\noindent\textbf{Corollary~{#2}}\;\textbf{#1}\  
}{\vskip 1em}

\newcommand{\Proofitem}[1]{\medskip \noindent $#1\;$} 
\newcommand{\Proofitemf}[1]{\noindent $#1\;$} 
\newcommand{\Defitem}[1]{\smallskip \noindent $#1\;$} 
 
\newcommand{\Defitemf}[1]{\noindent $#1\;$}

\newcommand{\hbra}{\noindent\hbox to \textwidth{\leaders\hrule height1.8mm depth-1.5mm\hfill}} 
\newcommand{\hket}{\noindent\hbox to \textwidth{\leaders\hrule height0.3mm\hfill}} 
\newcommand{\ratio}{.3}

\newtheorem{theorem}{Theorem} 
 
\newtheorem{definition}[theorem]{Definition} 
\newtheorem{lemma}[theorem]{Lemma} 
\newtheorem{corollary}[theorem]{Corollary} 
\newtheorem{proposition}[theorem]{Proposition}

\newtheorem{remark}[theorem]{Remark}

\newcommand{\Proof}{\noindent {\sc Proof}. } 
 
\newcommand{\qed}{\hfill${\Box}$}

\newcommand{\Figbar}{{\center \rule{\hsize}{0.3mm}}}    
 
\newcommand{\cl}[1]{{\cal #1}}          

\newcommand{\ol}[1]{\overline{#1}}      
 
\newcommand{\arrow}{\rightarrow}        
\newcommand{\Alt}{ \mid\!\!\mid  } 
\newcommand{\isum}{\oplus} 
 
\newcommand{\infer}[2]{\begin{array}{c} #1 \\ \hline #2 \end{array}} 
 
\newcommand{\Arrow}{\Rightarrow}        
 
\newcommand{\ubis}{\approx^u}          
\newcommand{\uabis}{\approx^{u}_{ccs}} 

\newcommand{\cbis}{\approx}        
\newcommand{\cdbis}{\approx_{\Uparrow}} 
\newcommand{\cabis}{\approx_{ccs}}  

\newcommand{\lcbis}{\approx^{\ell}} %
\newcommand{\lcdbis}{\approx^{\ell}_{\Uparrow}}

\newcommand{\lcbiswrong}{\approx^{\ell \Downarrow}} %

\newcommand{\union}{\cup}               
\newcommand{\inter}{\cap}               
\newcommand{\minus}{\backslash}         
\newcommand{\comp}{\circ}               
\newcommand{\set}[1]{\{#1\}}            
 
\newcommand{\dcl}{\downarrow}           
\newcommand{\conv}{\downarrow}          
\newcommand{\Conv}{\Downarrow}          
\newcommand{\CConv}{\Downarrow_{C}}
\newcommand{\Diver}{\Uparrow}           

\newcommand{\rel}[1]{{\cal #1}}         
 
\newcommand{\mand}{\mbox{ and }} 
\newcommand{\w}[1]{{\it #1}}    

\newcommand{\xst}[2]{\exists\, #1\;\: #2}

\newcommand{\s}[1]{{\sf #1}}    
\newcommand{\vc}[1]{{\bf #1}}

\newcommand{\act}[1]{\xrightarrow{#1}} 

 \newcommand{\wact}[1]{\stackrel{#1}{\Rightarrow}} 

\newcommand{\susp}{\downarrow} 
 
\newcommand{\wsusp}{\Downarrow}

 \newcommand{\pres}[2]{#1\triangleright #2} 
\newcommand{\present}[3]{{\sf present} \ #1 \ {\sf do } \ #2 \ {\sf  else} \ #3}

\newcommand{\tick}{{\sf tick}}          

\begin{document} 
 
\title{On convergence-sensitive bisimulation\\ and the
embedding of CCS in timed CCS
}
\author{Roberto M. Amadio\thanks{Work partially supported by ANR-06-SETI-010-02\
.} \\
        Universit\'e Paris Diderot\thanks{PPS, UMR-7126.}}

\maketitle

\begin{abstract} 
We propose a notion of convergence-sensitive bisimulation
that is built just over the notions of (internal) reduction and of
(static) context.
In the framework of timed CCS, we characterise this notion of 
`contextual' bisimulation via the usual labelled transition
system. 
We also remark that it provides  a suitable semantic framework 
for a fully abstract embedding of untimed processes into timed ones.
Finally, we show that the notion can be refined to
include sensitivity to divergence.
\end{abstract}

\section{Introduction}
The main motivation for this work is to build a notion of
convergence-sensitive bisimulation from first principles, namely from
the notions of {\em internal reduction} and of (static) {\em context}.  
A secondary motivation is to understand how asynchronous/untimed behaviours can be
embedded fully abstractly into synchronous/timed ones. 
Because the notion of convergence is very much connected to the notion of time, it 
seems that a convergence-sensitive bisimulation should find a 
natural application in a synchronous/timed context.  
Thus, in a nutshell, we are looking for an `intuitive' 
semantic framework that spans both
untimed/asynchronous and timed/synchronous models. 

For the sake of simplicity we will place our discussion 
in the well-known framework of (timed) CCS.
We assume the reader is familiar with CCS \cite{M89}.  Timed CCS
(TCCS) is a `timed' version of CCS whose basic principle is that {\em
time passes exactly when no internal computation is possible}. This
notion of `time' is inspired by early work on the {\sc Esterel}
synchronous language \cite{BC88}, and it has been formalised in
various dialects of CCS \cite{Yi91,NS94,HR95}.  Here we shall follow the
formalisation in \cite{HR95}.

As in CCS, one models the internal computation with an action $\tau$
while the passage of (discrete) time is represented by an action
$\tick$ that implicitly synchronizes all the processes and moves the
computation to the next instant.  \footnote{There seems to be no
standard terminology for this action. It is called $\epsilon$ in
\cite{Yi91}, $\chi$ in \cite{NS94}, $\sigma$ in \cite{HR95}, and
sometimes `next' in `synchronous' languages \`a la {\sc Esterel} 
\cite{A07}.}

In this framework, the basic principle we mentioned is formalised as
follows:
\[
P\act{\tick} \cdot \mbox{ iff } P \not \act{\tau} \cdot
\]
where we write $P\act{\mu}\cdot$ if $P$ can perform an action $\mu$.
TCCS is designed so that if $P$ is a process built with the usual CCS
operators and $P$ cannot perform $\tau$ actions then $P\act{\tick} P$. 
In other terms, CCS processes are {\em time insensitive}.
To compensate for this property, one introduces a new binary 
operator $\pres{P}{Q}$, called \w{else\_next}, that tries to run $P$ in the 
current instant and, if it fails, runs $Q$ in the following instant.

We assume countably many names $a,b,\ldots$ For each name $a$ there is
a communication action $a$ and a co-action $\ol{a}$.  We denote with
$\alpha,\beta,\ldots$ the usual CCS actions which are composed of
either an internal action $\tau$ or of a communication action
$a,\ol{a},\ldots$. We denote with $\mu,\mu',\ldots$ either an action
$\alpha$ or the distinct action $\tick$.

The TCCS processes $P,Q,\ldots$ are specified by the following grammar
\[
P::= 0 \Alt a.P \Alt P+P \Alt P\mid P \Alt \nu a \ P \Alt A(\vc{a}) \Alt \pres{P}{P}~.
\]
We denote with $\w{fn}(P)$ the names free in $P$.
We adopt the usual convention that for each thread identifier $A$ there is a
unique defining equation $A(\vc{b})=P$ where the parameters $\vc{b}$
include the names in $\w{fn}(P)$.
The related labelled transition system is specified in table \ref{lts}.

\begin{table}
{\footnotesize
\[
\begin{array}{cc}

\infer{}{a.P\act{a} P}

&\infer{P\act{a} P'\quad Q\act{\ol{a}}Q'}
{(P\mid Q)\act{\tau} (P'\mid Q')} \\ \\ 

\infer{P\act{\alpha}P'}
{(P\mid Q) \act{\alpha} (P'\mid Q)}

&\infer{P\act{\alpha}P'}
{P+Q\act{\alpha} P'} \\ \\ 

\infer{A(\vc{a})= P}
{A(\vc{b}) \act{\tau} [\vc{b}/\vc{a}]P} 

&\infer{P\act{\alpha} P'}
{\pres{P}{Q} \act{\alpha} P'} \\  \\

\infer{}
{0\act{\tick} 0}

&\infer{}
{a.P \act{\tick} a.P} \\ \\ 

\infer{P \not \act{\tau} \cdot}
{\pres{P}{Q} \act{\tick} Q}

&\infer{\begin{array}{c}
(P_1\mid P_2)\not\act{\tau}\cdot\\
P_i \act{\tick} Q_i\quad i=1,2
\end{array}}
{(P_1\mid P_2)\act{\tick} (Q_1\mid Q_2)} \\ \\

\infer{P_i\act{\tick}Q_i\quad i=1,2}
{P_1+P_2 \act{\tick} Q_1+Q_2} 

&\infer{P\act{\mu} Q \quad a,\ol{a} \neq \mu}
{\nu a \ P \act{\mu} \nu a \ Q}  

\end{array}
\]}
\caption{Labelled transition system}\label{lts}
\end{table}

Say that a process is a CCS process if it does not contain
the \w{else\_next} operator. The reader can easily verify that:

\Defitem{(1)} $P\act{\tick}\cdot$  if and only if $P\not\act{\tau} \cdot$.

\Defitem{(2)} If $P\act{\tick} Q_i$ for $i=1,2$ then $Q_1=Q_2$. One says that
the passage of time is {\em deterministic}.

\Defitem{(3)} If $P$ is a CCS process and $P\act{\tick}Q$ then $P=Q$. Hence
CCS processes are closed under labelled transitions. 
\\\smallskip
It will be convenient to write $\tau.P$ for $\nu a  \ (a.P \mid
\ol{a}.0)$ where $a\notin \w{fn}(P)$, $\tick.P$ for $\pres{0}{P}$,
and $\Omega$ for the diverging process $\tau.\tau.\ldots$.

\begin{remark}\label{lts-rmk}
\Defitemf{(1)} In the labelled transition system in table \ref{lts}, 
the definition of the $\tick$ action relies on the $\tau$ action and
the latter relies on the communication actions $a,a',\ldots$.
There is a well known method to give a direct definition of the
$\tau$ action that does not refer to the communication actions.
Namely, one defines (internal) reduction rules such as 
$(a.P+Q \mid \ol{a}.P'+Q') \arrow (P\mid P')$ which are applied modulo a suitable 
structural equivalence.

\Defitem{(2)} The labelled transition system in table \ref{lts} 
relies on {\em negative} conditions of the shape
$P\not\act{\tau}$. These conditions can be replaced by a condition
$\xst{L}{P\dcl L}$, where $L$ is a finite set of communication
actions. The predicate `$\dcl$' can be defined as follows:

{\footnotesize
\[
\begin{array}{ccc}
\infer{}
{0\dcl \emptyset}

&\infer{}{a.P\dcl \set{a}}

&\infer{P_i\dcl L_i,\quad i=1,2}
{(P_1+P_2) \dcl L_1\union L_2} \\  \\

\infer{P\dcl L}
{\pres{P}{Q}\dcl L} 

&\infer{P\dcl L}
{(\nu a \ P) \dcl L\minus\set{a,\ol{a}}}

&\infer{P_i\dcl L_i,\quad i=1,2\quad L_1\inter \ol{L_{2}}=\emptyset}
{(P_1\mid P_2) \dcl L_1\union L_2} 

\end{array}
\]}
\end{remark}

\subsection{Signals and a deterministic fragment}\label{sl-fragment}
As already mentioned, the TCCS model has been inspired by the notion of
time available in the {\sc Esterel} model \cite{BG92} 
and its relatives such as {\sc SL} \cite{BD95}.
These models rely on {\em signals} as the basic communication
mechanism. Unlike a channel, a signal persists within 
the instant and disappears at the end of it.
It turns out that a signal can be defined recursively in TCCS as:
\[
\s{emit}(a) = \pres{\ol{a}.\s{emit}(a)}{0}
\]
The `present' statement of {\sc SL} that either reads a signal and continues
the computation in the current instant or reacts to the absence
of the signal in the following instant can be coded as follows:
\[
\present{a}{P}{Q} = \pres{a.P}{Q}
\]
Modulo these encodings, the resulting fragment of TCCS is specified
as follows:
\[
P::= 0 \Alt \s{emit}(a) \Alt \present{a}{P}{P} \Alt (P\mid P) \Alt \nu
a \ P \Alt A(\vc{a})~.
\]
Notice that, unlike in (T)CCS, communication actions have an input or output polarity.
The most important property of this fragment is that its processes
are {\em deterministic} \cite{BD95,Amadio05}.

\subsection{The usual labelled bisimulation}\label{usual-bis}
As usual, one can define a notion of {\em weak transition} as follows:
\[
\wact{\mu} = \left\{
\begin{array}{ll}
(\act{\tau})^*  &\mbox{if }\mu = \tau \\
(\act{\tau})^* \comp \act{\mu} \comp (\act{\tau})^* 
&\mbox{otherwise} 
\end{array}\right.
\]
where the notation $X^*$ stands for the reflexive and transitive closure of
a binary relation $X$. When focusing just on internal reduction, we shall
write $\arrow$ for $\act{\tau}$ and $\Arrow$ for $\wact{\tau}$.
We write $P\arrow \cdot$ if
$\xst{P'}{(P\arrow P')}$, otherwise we say that $P$ has converged and
write $P\conv$.  We write $P\Conv$ if $\xst{Q}{(P\Arrow Q \mand
Q\conv)}$.  Thus $P\Conv$ means that $P$ {\em may} converge, {\em
i.e.}, there is a reduction sequence to a process that has converged.
Because $P\conv$ iff $P\act{\tick} \cdot$, we have that $P\Conv$ iff
$P \wact{\tick} \cdot$.

With respect to the notion of weak transition, one can define the
usual notion of bisimulation as the largest symmetric relation $\cl{R}$
such that if $(P,Q)\in \cl{R}$ and $P\wact{\mu} P'$ then
for some $Q'$, $Q\wact{\mu} Q'$ and $(P',Q')\in \cl{R}$. We denote with
$\ubis$ the largest labelled bisimulation ($u$ for {\em usual}).
When looking at CCS processes, one may focus on CCS actions
(thus excluding the $\tick$ action). We denote with $\uabis$ the
resulting labelled bisimulation.

\subsection{CCS vs. TCCS}
As we already noticed, TCCS has been designed so that CCS 
can be regarded as a transition closed subset of TCCS. 
A natural question is whether two CCS processes which are equivalent
with respect to an untimed environment are still equivalent
in a timed one. For instance, Milner  \cite{Milner83} discusses
a similar question when comparing CCS to SCCS.
\footnote{The notion of instant in SCCS is quite different from
the one considered in TCCS/{\sc Esterel}. In the former one declares
explicitly what each thread does at each instant while in the latter
the duration of an instant is the result of an arbitrarily complex 
interaction among the different threads.}

\subsubsection{Testing semantics}
In the context of TCCS and of a testing semantics, the question 
has been answered negatively by Hennessy and Regan \cite{HR95}.
For instance, they notice that the
processes $P = a.(b+c.b) + a.(d+c.d)$ and $Q = a.(b+c.d) + a.(d+c.b)$
are `untimed' testing equivalent but `timed' testing inequivalent.
The relevant test is the one that checks that if an action $b$ cannot follow
an action $a$ in the current instant then an action $b$ will happen in
the following instant just after an action $c$ (process $P$ will not
pass this test while process $Q$ does).  This remark motivated the
authors to develop a notion of `timed' testing semantics.

\subsubsection{Bisimulation semantics}
What is the situation with the usual labelled bisimulation semantics
recalled in section \ref{usual-bis}? Things are fine for {\em
reactive} processes which are defined as follows.

\begin{definition}
A process $P$ is reactive if whenever 
$P\wact{\mu_{1}} \cdots \wact{\mu_{n}} Q$, for $n\geq 0$,
we have the property that all sequences of $\tau$ reductions
starting from $Q$ terminate.
\end{definition}

\begin{proposition}\label{embed-react}
Suppose $P,Q$ are CCS reactive processes.
Then  $P\ubis Q$ if and only if $P\uabis Q$.
\end{proposition}
\Proof 
Clearly, $\ubis$ is a CCS bisimulation,  hence
$P\ubis Q$ implies $P\uabis Q$. To show the converse,
we prove that $\uabis$ is a timed bisimulation.
So suppose $P\uabis Q$ and $P\wact{\tick} P'$.
This means $P \wact{\tau} P_1 \act{\tick} P_1 \wact{\tau} P'$.
Then for some $Q_1$, $Q\wact{\tau} Q_1$ and $P_1\uabis Q_1$.
Further, because $Q_1$ is reactive there is a $Q_2$ such
that $Q_1\wact{\tau} Q_2$ and $Q_2 \conv$. 
By definition of bisimulation and the fact that $P_1\conv$, we
have that $P_1 \uabis Q_2$. 
So for some $Q'$, $Q_2 \wact{\tau} Q'$ and $P'\uabis Q'$.
Thus we have shown that there is a $Q'$ such that
$Q\wact{\tick} Q'$ and $P'\uabis Q'$. \qed \\

Proposition \ref{embed-react} fails when we look at {\em non-reactive} processes.
For instance, $0$ and $\Omega$ are regarded as untimed equivalent
but they are obviously timed inequivalent since the second process
does not allow time to pass.
This example suggests that if we want to extend proposition
\ref{embed-react} to non-reactive processes, then the notion of
bisimulation has to be {\em convergence sensitive}.

One possibility could be to adopt the usual bisimulation $\ubis$ on
CCS processes. We already noticed that if $P$ is a CCS process and
$P\act{tick} Q$ then $P=Q$. Thus in the bisimulation game between CCS
processes, the condition `$P\wact{\tick} P'$ implies $Q\wact{\tick} Q'$'
can be replaced by `$P\Conv$ implies $Q\Conv$'.  The resulting
equivalence on CCS processes is not new, for instance it appears in
\cite{LDH02} as the so called {\em stable} weak bisimulation.  
One may notice that this equivalence 
has reasonably good congruence properties.

\begin{proposition}\label{cong-usual-bis}
Suppose $P_1\ubis P_2$ and $Q_1\ubis Q_2$. Then

\Defitem{(1)} $(P_1 \mid R) \ubis (P_2 \mid R)$.

\Defitem{(2)} If $P_1,P_2\conv$ then 
$\pres{P_1}{Q_1} \ubis \pres{P_2}{Q_2}$.
\end{proposition}
\Proof First note that we can work with an asymmetric 
definition of bisimulation where a strong transition is matched by a weak one.

\Proofitemf{(1)} We just check the condition for the $\tick$ action.
Suppose $(P_1\mid R) \act{\tick} (P'_{1} \mid R')$.
This entails $P_1 \act{\tick} P'_1$ and $R\act{\tick} R'$.
Then $P_2 \wact{\tau} P''_2$, $P''_2 \conv$, and $P_1 \ubis P''_1$.
Also $P''_2 \wact{\tick} P'_2$ and $P'_1 \ubis P''_2$.
Finally, we have that $(P''_2 \mid R)\conv$ because if they
could synchronise on a name $a$ then so could $(P_1 \mid R)$.

\Proofitem{(2)}  There are two cases to consider.
If $\pres{P_1}{Q_1} \act{\tick} Q_1$ then
$\pres{P_{2}}{Q_{2}} \act{\tick} Q_2$.
If $\pres{P_{1}}{Q_{1}} \act{a} P'_{1}$ because
$P_{1}\act{a} P'_{1}$ then 
$P_{2} \wact{a}  P'_{2}$ and
$P'_{1} \ubis P'_{2}$. \qed

\begin{remark}
The \w{else\_next} operator suffers from the same
compositionality problems as the sum operator.  For instance, $0\ubis
\tau.0$ but $\pres{0}{Q} = \tick.Q$ while $\pres{\tau.0}{Q} \ubis 0$.
As for the sum operator, one may remark that in practice we are
interested in a {\em guarded} form of the \w{else\_next}
operator. Namely, the \w{else\_next} operator is only introduced as an
alternative to a communication action (the {\em present} operator
discussed in section \ref{sl-fragment} is such an
example). Proposition \ref{cong-usual-bis}(2) entails that in this form, 
the \w{else\_next} operator
preserves bisimulation equivalence.
\end{remark}

\subsubsection{An alternative path}
The reader might have noticed that on CCS processes $\ubis$ {\em refines}
$\uabis$ by adding may convergence as an observable along with the
usual labelled transitions. This is actually the case of all
convergence/divergence sensitive bisimulations we are aware of 
(see, {\em e.g.}, \cite{Walker90,LDH02}).
The question we wish to investigate 
is: what happens if we just take may convergence as an observable
without assuming the observability of the labelled transitions?
The question can be motivated by both pragmatic and mathematical 
considerations. On the pragmatic side, one may argue that
the normal operation of a timed/synchronous program is to receive
inputs at the beginning of each instant and to produce outputs at
the end of each instant. Thus, unless the instant terminates, no
observation is possible.  For instance, the process $(a\mid \Omega)$ 
could be regarded as equivalent to $\Omega$,  while they are
distinguished by the usual bisimulation $\ubis$ on the ground that the 
labelled transition $a$ is supposed to be directly observable.

On the mathematical side, it has been remarked by many authors that
the notion of labelled transition system is not necessarily
compelling. Specifically, one would like to  define a notion of
bisimulation without an {\em a priori} commitment to a notion of
label. To cope with this problem, a well-known approach started in
\cite{MS92} and elaborated in \cite{HY95} is to look at `internal'
reductions and at a basic notion of `barb' and then to close under
contexts thus producing a notion of `contextual'
bisimulation. However, even the notion `barb' is not always easy to
define and justify (an attempt based on the concept of {\em
bi-orthogonality} is described in \cite{RSS07}).  It seems to us that
a natural approach which applies to a wide variety of formalisms is to
regard convergence (may-termination) as the `intrinsic' basic
observable automatically provided by the internal reduction relation.

\subsubsection{Contribution}
Following these preliminary considerations, we are now in a position
to describe our contribution.
\begin{enumerate}

\item We introduce a notion of contextual bisimulation for CCS
whose basic observable (or barb) is the may-termination predicate 
(section \ref{convbis-sec}).

\item We provide various characterisations of this equivalence
culminating in one based on a suitable `convergence-sensitive'
labelled bisimulation (section \ref{char-sec}).

\item We derive from this characterisation that (section \ref{embed-sec}):
\begin{enumerate}

\item the embedding of CCS in TCCS is fully abstract (even for non-reactive
processes).

\item the proposed equivalence coincides with the usual one on reactive 
processes.

\item on non-reactive processes it identifies more processes than
the usual timed labelled bisimulation $\ubis$.

\item  on non-reactive CCS processes it is incomparable with 
the usual labelled CCS bisimulation $\uabis$.

\end{enumerate}

\item We refine the proposed notion of contextual bisimulation by making
it sensitive to {\em divergence} and show that the characterisation results mentioned
above can be extended to this case (section \ref{diver-sec}).

\end{enumerate}

The development will take place in the context of so called {\em weak}
bisimulation \cite{M89} which is more interesting and challenging than
{\em strong} bisimulation.

\section{Convergence sensitive bisimulation}\label{convbis-sec}
We denote with $C,D,\ldots$ one hole {\em static contexts} specified by
the following grammar:
\[
C::=[~] \Alt C \mid P \Alt \nu a C
\]
We require that the notion of bisimulation we consider
is preserved by the static contexts in the sense of \cite{HY95}.

\begin{definition}[bisimulation]\label{cbis-def}
A symmetric relation $\cl{R}$ on processes is a 
bisimulation if $P \rel{R} Q$ implies:
\begin{description}

\item[cxt]
for any static context $C$, $C[P] \rel{R} C[Q]$.

\item[red]
$P\wact{\mu} P'$, $\mu\in\set{\tau,\tick}$ implies 
$\xst{Q'}{(Q \wact{\mu} Q' \mbox{ and }P' \rel{R} Q')}$.

\end{description}
We denote with $\cbis$ the largest bisimulation. 
\end{definition}

\begin{remark}\label{CCS-tick-rmk}

\Defitemf{(1)} In view of remark \ref{lts-rmk}(1), the definition
\ref{cbis-def} of bisimulation does {\em not} assume the labels
$a,a',\ldots$ which correspond to the communication action.  Not only
the labels are not considered in the bisimulation game, but they are not
even required in the definition of the $\tau$ action. This means that
the definition can be directly transferred to more complex process
calculi where the definition of communication action is at best
unclear.

\Defitem{(2)}
For CCS processes, if $P\act{\tick} Q$ then $P=Q$.
It follows that in the definition above,
the condition {\bf [red]} when $\mu = \tick$ 
 can be replaced by $P\Conv$ implies $Q\Conv$.
This is obviously false for processes including
the \w{else\_next} operator; in this case one needs the $\tick$ 
action to observe the behaviour of processes after the
first instant, {\em e.g.}, to distinguish $\tick.a$ from $\tick.b$.

\end{remark}

In view of the previous remark, the definition of bisimulation is
specialised to CCS processes by simply restricting the condition {\bf
[cxt]} to CCS static contexts. We denote with $\cabis$ the
resulting largest bisimulation.

Next we remark that the observability of a 
`stable commitment (or barb)' is entailed by the observation
of convergence.

\begin{definition}
We say that $P$ (stably) commits on $a$, and write $P\Conv_{a}$, 
if $P \Arrow P'$, $P'\conv$, and $P'\act{a}\cdot$.\footnote{Note
that in this definition the process `commits' on action $a$ only when it has converged.}
\end{definition}

\begin{proposition}\label{bis-and-barb}
If $P \cbis Q$ and $P\Conv_{a}$ then $Q\Conv_{a}$.
\end{proposition}
\Proof 
Suppose $P\Conv_{a}$ and $P\cbis Q$.
Then $P\Arrow P'$, $P'\conv$, and $P'\act{a}\cdot$.
By definition of bisimulation, $Q\Arrow Q''$ and $P'\cbis Q''$.
Moreover, $Q''\Arrow Q'$, $Q'\conv$, $Q' \cbis P' \cbis Q''$.
To show that $Q'\act{a} \cdot$, consider the context
$C= ([~] \mid \ol{a}.\Omega)$. Then we have 
$C[P'] \not\Conv$, while $C[Q'] \Conv$ if and only if
$Q'\not \act{a} \cdot$. \qed \\

Another interesting notion is that of {\em contextual convergence}.

\begin{definition}
We say that a process $P$ is contextual convergent, and write
$P\CConv$, if  $\xst{C}{(C[P] \Conv)}$.
\end{definition}

Clearly, $P\Conv$ implies $P\CConv$ but the converse fails
taking, for instance, $(a+b) \mid \ol{a}.\Omega$.
Contextual convergence, can be characterised as follows.

\begin{proposition}\label{char-cconv}
The following conditions are equivalent:

\Defitem{(1)}
 $P\act{\alpha_{1}} \cdots \act{\alpha_{n}} P'$ and $P' \conv$.

\Defitem{(2)} There is a \w{CCS} \w{process} $Q$ such that $(P\mid Q) \Conv$.

\Defitem{(3)} $P\CConv$.
\end{proposition}

\Proof \Proofitemf{(1\Arrow 2)}
Suppose $P_0 \act{\alpha_{1}} P_1 \cdots \act{\alpha_{n}} P_n$ and
$P_n\susp$.
We build the process $Q$ in (2) by induction on $n$.
If $n=0$ we can take $Q=0$. Otherwise, suppose $n>0$.
By inductive hypothesis, there is $Q_1$ such that
$(P_1 \mid Q_1)\wsusp$. We proceed by case analysis on the
first action $\alpha_1$. 
If $\alpha_1=\tau$ take $Q=Q_1$ and if $\alpha_1=a$ 
take $Q=\ol{a}.Q_1$.

\Proofitem{(2\Arrow 3)}
Taking the static context $C=[~]\mid Q$.

\Proofitem{(3\Arrow 1)}
First, check by induction on a static context $C$
that $P\act{\tau}\cdot$ implies
$C[P]\act{\tau} \cdot$. Hence $C[P]\susp$ implies
$P\susp$.
Second, show that $C[P]\act{\alpha} Q$ implies
that $Q=C'[P']$ where $C'$ is a static context 
and either $P=P'$ or  $P\act{\alpha'}P'$.
Third, suppose 
$C[P] \act{\tau} Q_1 \cdots \act{\tau} Q_n$
with $Q_n\conv$. Show by induction on $n$ that
$P$ can make a series of labelled transitions
and reach a process which has converged. \qed \\

\begin{remark}\label{CCS-CConv-rmk}
As shown by the characterisation above, the notion
of contextual convergence is unchanged if we restrict
our attention to contexts composed of CCS processes.
\end{remark}

We notice that a bisimulation never identifies a process which is
contextual convergent with one which is not while identifying all
processes which are not contextual convergent.

\begin{proposition}\label{bis-and-cconv}
\Defitemf{(1)}  If $P\cbis Q$ and $P\CConv$ then $Q\CConv$.

\Defitem{(2)} If $P\not \CConv$ and $Q\not \CConv$ then $P\cbis Q$.

\end{proposition}
\Proof \Proofitemf{(1)} 
If $P\CConv$ then for some context $C$,
$C[P]\Conv$. By condition {\bf [cxt]}, we have
that $C[P] \cbis C[Q]$, and by condition
{\bf [red]} we derive that $C[Q]\Conv$. Hence $Q\CConv$.

\Proofitem{(2)}
We notice that the relation
$S=\set{(P,Q) \mid P,Q \not\CConv}$ is a bisimulation.
Indeed: 
(i) if $P\not \CConv$ then $C[P] \not \CConv$,
(ii) if $P \Arrow P'$ and $P\not\CConv$ then
$P'\not\CConv$, and
(iii) if $P\not \CConv$ then $P\not\wact{\tick} \cdot$. \qed

\section{Characterisation}\label{char-sec}
We characterise the (contextual and convergence sensitive)
bisimulation introduced in definition \ref{cbis-def} by means of a labelled
bisimulation. The latter is obtained from the former
by replacing condition {\bf [cxt]} with 
a suitable condition {\bf [lab]} 
on labelled transitions as defined in table \ref{lts}.

\begin{definition}[labelled bisimulation]\label{lcbis-def}
A symmetric relation $\cl{R}$ on processes is a 
labelled bisimulation if $P \rel{R} Q$ implies:
\begin{description}

\item[lab]
if $P\CConv$ and $P\wact{a} P'$ then $Q\wact{\alpha} Q'$ and
$P'\rel{R} Q'$ where $\alpha\in \set{a,\tau}$ and 
$\alpha = a$ if $P'\CConv$.

\item[red]
if $P\wact{\mu} P'$, $\mu\in\set{\tau,\tick}$ then 
$\xst{Q'}{(Q \wact{\mu} Q' \mbox{ and }P' \rel{R} Q')}$.

\end{description}
We denote with $\lcbis$ the largest labelled bisimulation. 
\end{definition}

\begin{remark}\label{lcbis-rmk}
\Defitemf{(1)}
By remark \ref{CCS-tick-rmk}, on CCS processes the condition
{\bf [red]} when $\mu=\tick$ is equivalent to: `$P\Conv$ implies
$Q\Conv$'. 
By remark \ref{CCS-CConv-rmk}, the notion of contextual convergence
is unaffected if we restrict our attention to CCS processes.
This means that, by definition, the (timed) labelled bisimulation restricted to
CCS processes is the same as the labelled bisimulation on (untimed) CCS
processes.

\Defitem{(2)}
The predicate of contextual convergence $\CConv$ plays an important
role in the condition {\bf [lab]}. To see why, 
suppose we replace it with the predicate $\Conv$ 
and assume we denote with $\lcbiswrong$ the
resulting largest labelled bisimulation.  The following example shows that
$\lcbiswrong$ is not preserved by parallel composition. Consider:
\[
\begin{array}{lll}
P_1 = a.(b+c),\quad
&P_2=a.b+a.c, \quad
&Q= \ol{a}.(d+\Omega)~.
\end{array}
\]
Then $(P_1 \mid Q) \lcbiswrong (P_2\mid Q)$ 
because both processes fail to converge. On the other hand, 
$(P_1\mid Q) \mid \ol{d} \not\lcbiswrong (P_2\mid Q)\mid \ol{d}$ 
because the first may converge to $(b+c)$ which cannot be matched
by the second process.

\Defitem{(3)}
One may consider an asymmetric and equivalent definition
of labelled bisimulation where strong transitions 
are matched by weak transitions. To check the equivalence,
it is useful to note that $P\not\CConv$ and $P\act{\alpha} P'$ implies
$P'\not\CConv$.

\end{remark}

We provide a rather standard 
proof that bisimulation and labelled bisimulation 
coincide.

\begin{proposition}\label{char-half1}
If $P\cbis Q$ then $P\lcbis Q$.
\end{proposition}
\Proof 
We show that the bisimulation $\cbis$ is a labelled bisimulation.
We denote with $P\isum Q$ the internal choice between
$P$ and $Q$ which is definable, {\em e.g.}, as $\tau.P+\tau.Q$.
Suppose $P\CConv$ and $P\wact{a} P'$. We consider
a context $C= [~]\mid T$ where $T= \ol{a}.((b\isum 0)\isum c)$ 
and $b,c$ are `fresh names' (not occurring in $P,Q$).
We know $C[P]\cbis C[Q]$ and $C[P] \Arrow (P'\mid (b\isum 0))$.
Thus $C[Q] \Arrow (Q'\mid T')$ where either 
$Q\wact{a} Q'$ and $T\wact{\ol{a}} T'$ or $Q\Arrow Q'$ and $T=T'$.

\Proofitem{\bullet}
Suppose $P'\not \CConv$. Then $(P'\mid (b\isum 0)) \not \CConv$ and,
by proposition \ref{bis-and-cconv}, $(Q'\mid T') \not \CConv$.
The latter implies that $Q'\not \CConv$. 
By contradiction, suppose $Q'\CConv$, that is $(Q'\mid R) \Conv$.
Then $(Q'\mid T') \mid R \mid \ol{T'} \Conv$ (contradiction!), where 
we take $\ol{T'} = \ol{a}$ if $T'=T$ and $\ol{T'}=0$ otherwise.
Hence, $P'\cbis Q'$ as required.

\Proofitem{\bullet} Suppose $P'\CConv$. 
If $Q\wact{a} Q'$ and $T\wact{\ol{a}} T'$ then we show that
it must be that $T'=(b\isum 0)$.
This is because if $P'\CConv$ then $P'\mid (b\isum 0) \CConv$ 
which in turn implies that for some $R$ (not containing the names $b$ or $c$), 
$(P'\mid (b\isum 0) \mid R) \Conv_{b}$. 
By proposition \ref{bis-and-barb}, we must have 
$Q''=(Q'\mid T')\mid R \Conv_{b}$. 
Thus $T'$ cannot be $0$ and it cannot be
$(b\isum 0)\isum c$, for otherwise 
$Q''\Conv_{c}$ which cannot be matched by $(P'\mid (b\isum 0) \mid R)$.
Further, we have $P'\mid (b\isum 0) \act{\tau} P' \mid 0 \ (= P')$.
So $(Q'\mid (b\isum 0)) \wact{\tau} (Q' \mid T'')$ and
$P'\cbis (Q'\mid T'')$. The latter entails that $T''=0$.

On the other hand, we show that $Q\wact{\tau} Q'$ and $T=T'$ is impossible.
Reasoning as above, we have $(P'\mid (b\isum 0) \mid R) \Conv_{b}$. 
But then if $(Q'\mid T)\mid R \Conv_{b}$ we shall also have
$(Q'\mid T)\mid R \Conv_{c}$. \qed \\

The following lemma relates contextual convergence to labelled
bisimulation (cf. the similar proposition \ref{bis-and-cconv}).

\begin{lemma}\label{cconv-and-lcbis}
\Defitemf{(1)}  If $P\lcbis Q$ and $P\CConv$ then $Q\CConv$.

\Defitem{(2)} If $P\not \CConv$ and $Q\not \CConv$ then $P\lcbis Q$.

\end{lemma}
\Proof
\Proofitemf{(1)}
By proposition \ref{char-cconv}, 
if $P\CConv$ then $P\act{\alpha_{1}} \cdots \act{\alpha_{n}} P'$ and
$P'\conv$. By definition of labelled bisimulation we should
have $Q\wact{\alpha_{1}} \cdots \wact{\alpha_{n}} Q'$ and $P'\lcbis Q'$.
Then  $P'\wact{\tick} \cdot$ entails
$Q'\wact{\tick}$, and therefore  $Q\CConv$.

\Proofitem{(2)}
By proposition \ref{bis-and-cconv}, 
$P,Q \not \CConv$ implies $P\cbis Q$, 
and by proposition \ref{char-half1} we conclude that
$P\lcbis Q$. \qed

\begin{proposition}\label{char-half2}
If $P\lcbis Q$ then $P\cbis Q$.
\end{proposition}
\Proof 
We show that labelled bisimulation is preserved by static contexts.
In view of remark \ref{lcbis-rmk}(3), we shall work
with an asymmetric definition of bisimulation. With respect to this
definition,  we show that the following relations are labelled bisimulations:
\[
\set{(\nu a \ P, \nu a\ Q) \mid P\lcbis Q} \union \lcbis \ ,  \qquad
\set{(P\mid R,Q\mid R) \mid P\lcbis Q} \union \lcbis ~.
\]
The case for restriction is a routine verification so we
focus on parallel composition.
Suppose $(P\mid R) \act{\mu} \cdot$. We proceed by case analysis.

\Proofitem{\bullet} 
$(P\mid R) \act{\alpha} (P\mid R')$ because 
$R\act{\alpha} R'$.
Then $(Q\mid R) \act{\alpha} (Q\mid R')$.

\Proofitem{\bullet}
$(P\mid R) \act{\tick} (P'\mid R')$ because
$P\act{\tick} P'$ and $R\act{\tick} R'$.
Then $Q\Arrow Q_1 \act{\tick} Q_2 \Arrow Q'$ and $P'\lcbis Q'$. 
Notice that $P \lcbis Q_1$ with $P,Q_1 \conv$, and therefore 
$(Q_1 \mid R) \act{\tick} (Q_2 \mid R')$.
Hence $(Q\mid R) \wact{\tick} (Q'\mid R')$.

\Proofitem{\bullet}
Suppose  $(P\mid R) \CConv$ and
$(P\mid R) \act{a} (P'\mid R)$ because
$P\act{a} P'$.  Then $P \CConv$ and therefore
$Q\wact{\alpha} Q'$, $\alpha \in \set{a,\tau}$,
and $P'\lcbis Q'$. 
If $(P'\mid R)\CConv$ then 
$P' \CConv$ and this entails
$\alpha =a$.

\Proofitem{\bullet}
Suppose $(P\mid R) \act{\tau} (P'\mid R)$ because
$P\act{\tau} P'$. Then $Q\wact{\tau} Q'$ and
$P'\lcbis Q'$.

\Proofitem{\bullet}
Suppose $(P\mid R)\act{\tau} (P'\mid R')$ because
$P\act{a} P'$ and $R\act{\ol{a}} R'$. 
If $P, P'\CConv$ then $Q\wact{a} Q'$ and $P'\lcbis Q'$.
If $P\CConv$ and $P'\not\CConv$ then $Q\wact{\alpha} Q'$,
$\alpha\in\set{a,\tau}$, and $P'\lcbis Q'$.
But then $(P'\mid R), (Q'\mid R) \not \CConv$, 
and we apply lemma \ref{cconv-and-lcbis}.
If $P\not \CConv$ then $Q\not\CConv$ and therefore 
$(Q\mid R) \not \CConv$, and we apply again lemma
\ref{cconv-and-lcbis}. \qed  \\

As a first application of the characterisation we check that
bisimulation is preserved by the {\em else\_next} operator
in the sense of proposition \ref{cong-usual-bis}(2).

\begin{corollary}
Suppose $P_1\cbis P_2$, $P_1,P_2\conv$, and $Q_1\cbis Q_2$. Then
$\pres{P_1}{Q_1} \cbis \pres{P_2}{Q_2}$.
\end{corollary}
\Proof There are two cases to consider.
If $\pres{P_1}{Q_1} \act{\tick} Q_1$ then
$\pres{P_{2}}{Q_{2}} \act{\tick} Q_2$.
If $\pres{P_{1}}{Q_{1}} \act{a} P'_{1}$ because
$P_{1}\act{a} P'_{1}$ then 
$P_{2} \wact{\alpha}  P'_{2}$,
$P'_{1} \lcbis P'_{2}$, and $\alpha\in\set{\tau,a}$.
We note that it must be that $\alpha = a$. 
Indeed, if $\alpha=\tau$ then since $P_{2}\conv$ we
must have $P'_{2}=P_{2}$ and $P'_{1} \CConv$. 
The latter forces $\alpha=a$ which is a contradiction. \qed

\section{Embedding CCS in TCCS}\label{embed-sec}
In this section we collect some easy corollaries of the characterisation.
First, we remark that two CCS processes 
are bisimilar when observed in an untimed/asynchronous
environment if and only if they are bisimilar in a 
timed/synchronous environment.

\begin{proposition}
Suppose $P,Q$ are CCS processes. Then
$P\cbis Q$ if and only if $P\cabis Q$.
\end{proposition}
\Proof  
By propositions \ref{char-half1} and \ref{char-half2} we
know that $\cbis = \lcbis$. 
By remark \ref{lcbis-rmk}(1), the labelled bisimulation on untimed
processes coincides with the restriction to CCS processes 
of the timed labelled bisimulation. \qed \\

Second, we compare the notion of convergence-sensitive bisimulation
we have introduced with the usual one we have recalled
in the section \ref{usual-bis}. All the notions collapse
on reactive processes.

\begin{proposition}
Suppose $P,Q$ are reactive processes. Then
$P\cbis Q$ if and only if $P\ubis Q$.
\end{proposition}
\Proof  
We know that $\cbis = \lcbis$.
Reactive processes are
closed under labelled transitions and on reactive processes
the conditions that define
labelled bisimulation coincide with the ones for
the usual bisimulation. \qed \\

The situation on non-reactive processes is summarised as follows
where all implications are strict.

\begin{proposition}
Suppose $P,Q$ are processes.

\Defitem{(1)}
If $P\ubis Q$ then $P\cbis Q$.

\Defitem{(2)} 
If moreover $P$ and $Q$ are CCS processes then
$P\ubis Q$ implies both $P\uabis Q$ and $P\cbis Q$.

\end{proposition}
\Proof
\Proofitemf{(1)}
The clauses in the definition of $\ubis$ imply directly
those in the definition of the labelled bisimulation that
characterises $\cbis$ (definition \ref{lcbis-def}).
To see that the converse fails note that 
$(a\mid \Omega) \cbis \Omega$ while $(a\mid \Omega) \not\ubis \Omega$.

\Proofitem{(2)} 
Use (1) and the fact that the clauses in the definition of $\ubis$ 
imply directly those in the definition of
$\uabis$. To see that the converse fails
use the counter-example in (1) and  the fact that 
$0 \uabis \Omega$ while $0\not\ubis \Omega$. \qed

\section{Divergence sensitive bisimulation}\label{diver-sec}
We refine the notion of bisimulation to make it sensitive
to {\em divergence} and show that the characterisation 
presented in section \ref{char-sec} can be adapted to this case.

We say that a process $P$ may diverge and write $P\Diver$ 
if there is an infinite reduction
sequence of $\tau$ actions that starts from $P$.
We refine the notion of bisimulation by making it sensitive to
divergence.

\begin{definition}[$\Diver$-bisimulation]\label{diverbis-def}
A symmetric relation $\cl{R}$ on processes is a 
divergence sensitive bisimulation ($\Diver$-bisimulation, for short)
if it is a bisimulation according to definition 
\ref{cbis-def} and if $P \rel{R} Q$ and $P\Diver$ implies 
$Q\Diver$. We denote with $\cdbis$ the largest $\Diver$-bisimulation.
\end{definition}

\begin{remark}
Say that a process $P$ is strongly normalising if all reduction
sequences of $\tau$-actions that start from $P$ terminate.  A process
is strongly normalising if and only if it may not diverge.  It follows
that one can give an equivalent formulation of $\Diver$-bisimulation
by replacing the may divergence predicate with the strong
normalisation predicate.
\end{remark}

We notice the following properties whose proof is direct.

\begin{proposition}\label{easy-diver-facts}
\Defitemf{(1)} 
If $P\cdbis Q$ then $P\cbis Q$.

\Defitem{(2)} If $P\cdbis Q$ and $P \Conv_{a}$ then $Q\Conv_{a}$.

\Defitem{(3)} If $P\cdbis Q$ and $P\CConv$ then $Q\CConv$.

\Defitem{(4)} If $P\not\CConv$ then $P\Diver$.

\Defitem{(5)} If $P\not\CConv$ and $Q\not\CConv$ then $P\cdbis Q$.
\end{proposition}

\Proof
\Proofitemf{(1)} 
A $\Diver$-bisimulation is also a bisimulation.

\Proofitem{(2)}
We apply (1) and proposition \ref{bis-and-barb}.

\Proofitem{(3)}
We apply (1) and proposition \ref{bis-and-cconv}(1).

\Proofitem{(4)} Immediate, by definition.

\Proofitem{(5)} If $P\not\CConv$ and $Q\not\CConv$
then $P\Diver$ and $Q\Diver$. \qed \\

It follows that $\Diver$-bisimulation coincides with bisimulation
on the processes that are not contextual convergent. On the other
hand, on those that are contextual convergent, it is a strictly finer
notion as, {\em e.g.}, it distinguishes $0$ from $A=\tau.A+\tau.0$.

The characterisation of $\Diver$-bisimulation turns out to be
straightforward: it is enough to make the labelled bisimulation
we have introduced in definition \ref{lcbis-def} sensitive to divergence.

\begin{definition}[$\Diver$-labelled bisimulation]\label{lcdbis-def}
A symmetric relation $\cl{R}$ on processes is a divergence sensitive labelled
bisimulation (or $\Diver$-labelled bisimulation) if 
it is a labelled bisimulation and if $P \rel{R} Q$ 
and $P\Diver$ implies that $Q\Diver$. 
We denote with $\lcdbis$ the largest $\Diver$-labelled bisimulation. 
\end{definition}

Because of the properties stated in proposition
\ref{easy-diver-facts}, 
one can repeat the proofs in section \ref{char-sec}
while adding specific arguments to take the sensitivity
to divergence into account.

\begin{proposition}\label{dchar-half1}
If $P\cdbis Q$ then $P\lcdbis Q$.
\end{proposition}
\Proof 
We show that $\cdbis$ is a $\Diver$-labelled bisimulation by repeating
the proof schema in  proposition \ref{char-half1}. 
Note that the condition that refers to divergence is the same
for $\Diver$-bisimulation and for $\Diver$-labelled bisimulation. \qed

\begin{lemma}\label{cconv-and-lcdbis}
\Defitemf{(1)}  If $P\lcdbis Q$ and $P\CConv$ then $Q\CConv$.

\Defitem{(2)} If $P\not \CConv$ and $Q\not \CConv$ then $P\lcdbis Q$.
\end{lemma}
\Proof 
\Proofitemf{(1)}
Note that $P\lcdbis Q$ implies $P\lcbis Q$ and apply 
lemma \ref{cconv-and-lcbis}(1).

\Proofitem{(2)}
By proposition \ref{easy-diver-facts}(5), 
$P\not\CConv$ and $Q\not\CConv$ implies $P\cdbis Q$ and 
by proposition \ref{dchar-half1} the latter implies 
$P\lcdbis Q$.  \qed

\begin{proposition}\label{dchar-half2}
If $P\lcdbis Q$ then $P\cdbis Q$.
\end{proposition}
\Proof 
As in proposition \ref{char-half2}, we have to verify that $\lcdbis$ 
is preserved by name generation and parallel composition.
For the former we note that $\nu a \ P \Diver$ if and only if $P \Diver$.
For the latter, we can repeat the proof in proposition \ref{char-half2}.
Moreover, we have to consider the case where 
$P\lcdbis Q$ and $(P\mid R) \Diver$.
The process $(P\mid R)$ diverges because: 
either $P$ and $R$ may engage in a finite
number of synchronisations after which one of the two diverges 
or  $P$ and $R$ may engage in an infinite
number of synchronisations. 
Suppose the finite or infinite number of synchronisations between 
$P$ and $R$ correspond to  the transitions 
$P\wact{a_{1}} P_1 \wact{a_{2}} \cdots$ and
$R\wact{\ol{a}_{1}} R_1 \wact{\ol{a}_{2}} \cdots$
If $P,P_1,\cdots$ are all contextually convergent then
$Q\wact{a_{1}} Q_1 \wact{a_{2}} \cdots$ and
$P_i \lcdbis Q_i$. Hence $(Q\mid R) \Diver$.
If $P\not \CConv$ then $Q\not\CConv$ implies $(Q\mid R)\not\CConv$ 
which implies $(Q\mid R)\Diver$.
Finally, suppose $P_i$ is the least $i$ such that $P_i\not\CConv$.
Then $Q\wact{a_{1}} \cdots \wact{a_{i-1}} Q_{i-1} \wact{\alpha_{i}} Q_i$
with $Q_i\not\CConv$ and $\alpha_{i}\in \set{a_i,\tau}$.
If $\alpha_{i}=a_{i}$ then $(Q\mid R) \Diver$ because
$(Q\mid R) \wact{\tau} (Q_i\mid R')$ and $Q_i \Diver$.
If $\alpha_{i}=\tau$ then $(Q\mid R)\Diver$ because
$(Q\mid R) \wact{\tau} (Q_{i} \mid R_{i-1})$ and $Q_i \Diver$. \qed

\section{Conclusion}
We have presented a natural notion of contextual and convergence
sensitive bisimulation and we have shown that it can be characterised
by a variant of the usual notion of labelled bisimulation relying on
the concept of contextual convergence. As a direct corollary of this
characterisation, we have shown that (untimed) CCS processes are
embedded fully abstractly into timed ones. Finally, we have refined
the notion of bisimulation to make it divergence-sensitive.

We believe that our main contribution, if any, is of a methodological
nature. The notion of bisimulation we have introduced just requires the
notions of reduction and static context as opposed to previous
approaches that build on the notion of `labelled' transition or on the
notion of `barb'.  It would be interesting to apply the proposed
approach to other situations where the notion of equivalence is
unclear.
For instance, we expect that our results can be extended to a TCCS
with `asynchronous' communication or with `signal-based'
communication.

\end{document}